\documentclass[11pt]{article}
\usepackage{times}
\usepackage{amsmath,amsthm}
\usepackage{color,setspace,epsfig,makecell}
\usepackage[numbers,sort&compress,round]{natbib}
\usepackage[top=3cm,left=3.3cm,right=3.3cm,bottom=3.5cm]{geometry}

\makeatletter
  \def\@eqnnum{{\normalfont \normalcolor [\theequation]}}
\makeatother

\makeatletter
  \def\tagform@#1{\maketag@@@{[#1]\@@italiccorr}}
\makeatother

\makeatletter
\renewcommand\tableofcontents{\@starttoc{toc}}
\makeatother

\newtheoremstyle{plain}
{12pt}
{9pt}
{}
{}
{\bfseries}
{}
{\newline}
{}

\newtheoremstyle{lemma}
{12pt}
{9pt}
{\normal }
{}
{\bfseries}
{.}
{.3cm}
{}

\theoremstyle{plain}
\newtheorem{Prop}{Proposition}

\theoremstyle{lemma}

\title{\sffamily\textbf{Zero-determinant alliances in\\ multiplayer social dilemmas}}
\author{\parbox[c]{12.5cm}{\center{Christian Hilbe$^1$, Arne Traulsen$^2$, Bin Wu$^2$ \& Martin A. Nowak$^{1,3}$}\\[0.2cm]
\normalsize $^1$ Program for Evolutionary Dynamics, Harvard University, Cambridge, MA~02138, USA\\[0.1cm]
$^2$ EvolutionaryTheory Group, Max-Planck-Institute for Evolutionary Biology, August-Thienemann-Stra{\ss}e 2, 24306 Pl\"{o}n, Germany\\[0.1cm]
$^3$ {Department of Organismic and Evolutionary Biology, Department of Mathematics, Harvard University, Cambridge, MA 02138}}}
\date{\today}

\begin{document}
\maketitle

\onehalfspacing
\noindent {\bf Direct reciprocity and conditional cooperation are important mechanisms to prevent free riding in social dilemmas. But in large groups these mechanisms may become ineffective, because they require single individuals to have a substantial influence on their peers. However, the recent discovery of the powerful class of zero-determinant strategies in the iterated prisoner's dilemma suggests that we may have underestimated the degree of control that a single player can exert. Here, we develop a theory for zero-determinant strategies for multiplayer social dilemmas, with any number of involved players. We distinguish several particularly interesting subclasses of strategies: fair strategies ensure that the own payoff matches the average payoff of the group; extortionate strategies allow a player to perform above average; and generous strategies let a player perform below average. We use this theory to explore how individuals can enhance their strategic options by forming alliances.  The effects of an alliance depend on  the size of the alliance, the type of the social dilemma, and on the strategy of the allies: fair alliances reduce the inequality within their group; extortionate alliances outperform the remaining group members;  but generous alliances increase welfare. Our results highlight the critical interplay of individual control and alliance formation to succeed in large groups.\\}

\newpage

\noindent
Cooperation among self-interested individuals is generally difficult to achieve \cite{hardin:Science:1968,olson:book:1971}, but typically the free rider problem is aggravated even further when groups become large \cite{boyd:JTB:1988, hauert:prsb:1997, suzuki:JTB:2007a,santos:pnas:2011, abouchakra:PLoSCB:2012,Yang:PNAS:2013}. In small communities, cooperation can often be stabilized by forms of direct and indirect reciprocity \cite{trivers:QRB:1971,axelrod:book:1984, alexander:book:1987,wedekind:Science:2000, henrich:AER:2001a,nowak:Nature:2005, van-veelen:PNAS:2012,zagorsky:plosone:2013}. For large groups, however, it has been suggested that these mechanisms may turn out to be ineffective, as it becomes more difficult to keep track of the reputation of others, and because the individual influence on others diminishes \cite{boyd:JTB:1988, hauert:prsb:1997, suzuki:JTB:2007a,santos:pnas:2011,abouchakra:PLoSCB:2012}. To prevent the tragedy of the commons, and to compensate for the lack of individual control, many successful communities have thus established central institutions that enforce mutual cooperation \cite{yamagishi:JPSP:1986,ostrom:book:1990,sigmund:nature:2010, Guala:BBS:2012}. 

However, a recent discovery suggests that we may have underestimated the amount of control that single players can exert in repeated games. For the repeated prisoner's dilemma, Press and Dyson \cite{press:pnas:2012} have shown the existence of {\it zero-determinant strategies} (or ZD strategies), which allow a player to unilaterally enforce a linear relationship between the own payoff and the co-player's payoff  -- irrespective of the co-player's actual strategy. The class of zero-determinant strategies is surprisingly rich: For example, a player who wants to ensure that the own payoff will always match the co-player's payoff can do so by applying a {\it fair} ZD strategy, like Tit-for-Tat. On the other hand, a player who wants to outperform the respective opponent can do so by slightly tweaking the Tit-for-Tat strategy to the own advantage, thereby giving rise to {\it extortionate} ZD strategies. The discovery of such strategies has prompted several theoretical studies, exploring how different ZD strategies evolve under various evolutionary conditions \cite{stewart:pnas:2012,ball:nature:2012,hilbe:pnas:2013,akin:2013,stewart:pnas:2013,hilbe:plosone:2013b,szolnoki:pre:2014}. 

However, ZD strategies are not confined to pairwise games. In a recently published study, it was shown that ZD strategies also exist in repeated public goods games \cite{Pan:arxiv:2014}; and herein we will show that such strategies exist for {\it all} symmetric social dilemmas, with an arbitrary number of participants. We also use this theory to demonstrate how zero-determinant strategists can even further enhance their strategic options by forming alliances. As we will show, the impact of an alliance depends on its size, the type of the social dilemma and on the applied strategy of the allies: while fair alliances reduce inequality within their group, extortionate alliances strive for unilateral advantages, with larger alliances being able to enforce even more extortionate relationships. These results suggest that when a single individual's strategic options are limited, forming an alliance may result in a considerable leverage. 

\begin{figure}[t]
\centering
\includegraphics[width=10cm]{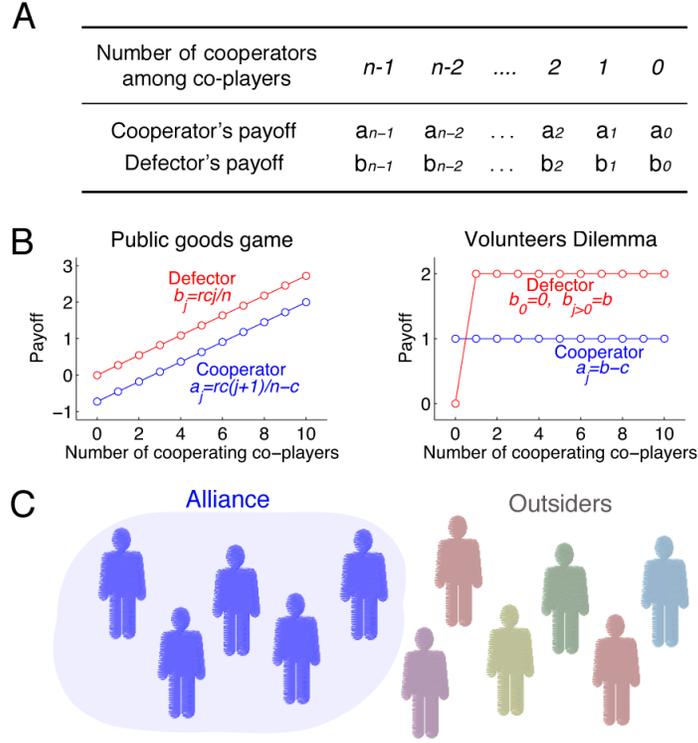}
\caption{Illustration of the model assumptions  for repeated social dilemmas. (A)~We consider symmetric social dilemmas in which each player can either cooperate or defect. The player's payoff depends on the own decision, and on the number of other group members who decide to cooperate. 
(B) We will discuss two particular examples: the public goods game (in which payoffs are proportional to the number of cooperators), and the volunteers dilemma (as the most simple example of a nonlinear social dilemma).
(C) Alliances are defined as a collection of individuals who coordinate on a joint ZD strategy. We refer to the set of group members that are not part of the alliance as outsiders. Outsiders are not restricted to any particular strategy; in particular, they may choose a joint strategy themselves.} \label{Fig:Illustration1}
\end{figure}

To obtain these results, we consider a repeated social dilemma between $n$ players. In each round of the game players can independently decide whether to cooperate (C) or to defect (D). A player's payoff depends on the player's own decision, and on the decisions of all other group members (see Fig.~\ref{Fig:Illustration1}A): in a group in which $j$ of the other group members cooperate, a cooperator receives the payoff $a_j$, whereas a defector obtains $b_j$. We assume that payoffs satisfy the following three properties that are characteristic for social dilemmas (corresponding to the `individual-centered' interpretation of altruism in \cite{kerr:tree:2004}): ({\it i}) Irrespective of the own strategy, players prefer the other group members to cooperate ($a_{j+1}\ge a_j$ and $b_{j+1}\ge b_j$ for all $j$). ({\it ii}) Within any mixed group, defectors obtain strictly higher payoffs than cooperators ($b_{j+1} > a_j$ for all $j$). ({\it iii}) Mutual cooperation is favored over mutual defection ($a_{n-1}>b_0$). To illustrate our results, we will discuss two particular examples of multiplayer games (see Fig.~\ref{Fig:Illustration1}B). In the first example, the public goods game \cite{ledyard:bookchapter:1995}, cooperators contribute an amount $c>0$ to a common pool, knowing that total contributions are multiplied by $r$ (with $1<r<n$) and evenly shared among all group members. Thus, a cooperator's payoff is $a_j=rc(j+1)/n-c$, whereas defectors yield $b_j=rcj/n$. In the second example, the volunteer's dilemma \cite{diekmann:jcr:1985}, at least one group member has to volunteer to bear a cost $c>0$ in order for all group members to derive a benefit $b>c$. Therefore, cooperators obtain  $a_j=b-c$ (irrespective of $j$) while defectors yield $b_j=b$ if $j\ge 1$ and $b_0=0$. Both examples (and many more, such as the collective-risk dilemma,  \cite{milinski:pnas:2008,santos:pnas:2011,abouchakra:PLoSCB:2012}) are simple instances of multiplayer social dilemmas.

We assume that the social dilemma is repeated, such that individuals can react to their co-players' past actions (for simplicity, we will focus here on the case of an infinitely repeated game, but our results can easily be extended to the finite case, see Supporting Information). As usual, payoffs for the repeated game are defined as the average payoff that players obtain over all rounds. In general, strategies for such repeated games can become arbitrarily complex, as subjects may condition their behavior on past events and on the round number in non-trivial ways. Nevertheless, as in pairwise games \cite{press:pnas:2012}, zero-determinant strategies are surprisingly simple -- they depend on the outcome of the last round only. In contrast to most previous studies on repeated games, however, we do not presume that individuals act in isolation. Instead, we allow subjects to form an alliance, which can be considered as a collection of players who coordinate on a joint strategy (see Fig.~\ref{Fig:Illustration1}C). In the following, we will discuss how the strategic power of such an alliance depends on the number of involved players, on the social dilemma, and on the applied ZD strategy of the allies. 

\section*{Results}

\noindent {\bf Zero-determinant strategies in large-scale social dilemmas.} ZD strategies are particular memory-one strategies \cite{nowak:nature:1993,hauert:prsb:1997,sigmund:book:2010,press:pnas:2012}; they only condition their behavior on the outcome of the previous round.  Memory-one strategies can be written as a vector $(p_{S,j})$, where $p_{S,j}$ denotes the probability to cooperate in the next round, given that the player previously played $S\in\{C,D\}$, and that $j$ of the co-players cooperated. ZD strategies have a particular form (see also Supporting Information): players with a ZD strategy set their cooperation probabilities such that
\begin{equation} \label{Eq:DefZD}
\begin{array}{llc}
p_{C,j}	&=	&\displaystyle 1+\phi \Big[(1-s)(l-a_j)-\frac{n-j-1}{n-1}(b_{j+1}-a_j)\Big]\\[0.25cm]
p_{D,j} 	&=	&\displaystyle ~\phi \Big[(1-s)(l-b_j)+\frac{j}{n-1}(b_j-a_{j-1})\Big],
\end{array}
\end{equation}
where $a_j$ and $b_j$ are the specific payoffs of the social dilemma (as outlined in Fig.~1), and where $l$, $s$, and $\phi>0$ are parameters that can be chosen by the player. While these ZD strategies may appear inconspicuous, they give players an unexpected control over the resulting payoffs of the game, as we will show below.

Instead of presuming that players act in isolation, as in previous models of zero-determinant strategies \cite{press:pnas:2012,stewart:pnas:2012,ball:nature:2012,hilbe:pnas:2013,akin:2013,stewart:pnas:2013,hilbe:plosone:2013b,szolnoki:pre:2014,Pan:arxiv:2014} we explicitly allow subjects to form alliances, and to coordinate on some joint ZD strategy. Let $k$ be the number of allies, with $1\le k<n-1$ (in particular, this covers the case $k=1$ of solitary alliances). In the Supporting Information we prove that if each of the allies applies the same ZD-strategy with parameters $l$, $s$, and $\phi$, then payoffs satisfy the equation 
\begin{equation} \label{Eq:PropZD}
\pi_{-\mathcal{A}}=s_\mathcal{A}\pi_\mathcal{A}+(1-s_\mathcal{A})l,
\end{equation}
where $\pi_\mathcal{A}$ is the mean payoff of the allies, $\pi_{-\mathcal{A}}$ is the mean payoff of all outsiders, and 
\begin{equation}
 \label{Eq:sEff}
s_\mathcal{A}= \frac{s(n-1)-(k-1)}{n-k}. 
\end{equation}
Relation [\ref{Eq:PropZD}] suggests that by using a ZD strategy, alliances exert a direct influence on the payoffs of the outsiders. This relation is remarkably general, as it is independent of the specific social dilemma being played, and as it is fulfilled irrespective of the strategies that are adopted by the outsiders (in particular, outsiders are not restricted to memory-one strategies; it even holds if some or all of the outsiders coordinate on a joint strategy themselves). We call the parameter  $l$ the baseline payoff, $s$ the slope of the applied ZD strategy, and $s_\mathcal{A}$ the effective slope of the alliance. In the special case of a single player forming an alliance, $k=1$, the effective slope of the alliance according to Eq. [\ref{Eq:sEff}] simplifies to $s_\mathcal{A}=s$. 

The parameters $l,s,$ and $\phi$ of a ZD strategy cannot be chosen independently, as the entries $p_{S,j}$ of a ZD strategy are probabilities that need to satisfy  $0\le p_{S,j}\le 1$. In general, the admissible parameters depend on the specific social dilemma being played. In the Supporting Information we show that exactly those relations [\ref{Eq:PropZD}] can be enforced for which either $s_\mathcal{A}=s=1$ (in which case the parameter $l$ in the definition of ZD strategies is irrelevant), or for which the parameters $l$ and $s_\mathcal{A}$ satisfy
\begin{equation} \label{Eq:CharPayRelAlliances}
\displaystyle
\begin{array}{ccccc}
\max \!\left\{b_j-\frac{j}{(1-s_\mathcal{A})(n-k)}(b_j-a_{j-1})\right\} \!\!  &\!\!  \le \!\!\! 	&\!\! l \!\! 	&\!\!\!  \le \!\!  &\!\! \min \! \left\{a_j+\frac{n-j-1}{(1-s_\mathcal{A})(n-k)}(b_{j+1}-a_j)\right\},\\
\end{array}
\end{equation}
where the maximum and minimum is taken over all possible group compositions, $0\le j \le n-1$.
It follows that feasible baseline payoffs are bounded by the payoffs for mutual cooperation and mutual defection, $b_0\le l\le a_{n-1}$, and that the effective slope needs to satisfy the inequality $-1/(n-k)\le s_\mathcal{A}\le 1$. Moreover, as social dilemmas satisfy $b_{j+1}>a_j$ for all $j$, condition [\ref{Eq:CharPayRelAlliances}] implies that the range of enforceable payoff relations is strictly increasing with the size of the alliance -- larger alliances are able to enforce more extreme relationships between the payoffs of the allies and the outsiders. In the following, we will highlight several special cases of ZD strategies, and we discuss how subjects can increase their strategic power by forming alliances.\\ 

\noindent {\bf Fair alliances.} As a first example, let us consider alliances that apply a ZD strategy with slope $s_\mathcal{A}=s=1$. By Eq.~[\ref{Eq:PropZD}], such alliances enforce the payoff relation $\pi_\mathcal{A}=\pi_{-\mathcal{A}}$, such that the allies' mean payoff matches the mean payoff of the outsiders. We call such ZD strategies (and the alliances that apply them) {\it fair}. As shown in Figure 2A, fair strategies do not ensure that all group members get the same payoff -- due to our definition of social dilemmas,  unconditional defectors always outperform unconditional cooperators, no matter whether the group also contains fair players. Instead, fair players can only ensure that they do not take any unilateral advantage of their peers.  Interestingly, it follows from Eq.~[\ref{Eq:sEff}] that fair alliances consist of fair players: because $s_\mathcal{A}=1$ implies $s=1$, each player $i$ of a fair alliance individually enforces the relation $\pi_i=\pi_{-i}$. It also follows from our characterization [\ref{Eq:CharPayRelAlliances}] that such fair ZD strategies exist for all social dilemmas -- irrespective of the particular payoffs and irrespective of the group size. 

As an example, let us consider the strategy {\it proportional Tit-for-Tat} ($pTFT$), for which the  probability to cooperate is simply given by the fraction of cooperators among the co-players in the previous round, 
\begin{equation}\label{eq:ptft}
pTFT_{S,j}=\frac{j}{n-1}.
\end{equation}
For pairwise games, this definition of $pTFT$ simplifies to the classical Tit-for-Tat strategy. However, also for the public goods game and for the volunteer's dilemma, $pTFT$ is a ZD strategy (because it can be obtained from Eq.~[\ref{Eq:DefZD}] by setting $s=1$ and $\phi=1/c$, with $c$ being the cost of cooperation). As $s=1$, alliances of $pTFT$ players are fair, and they enforce $\pi_\mathcal{A}=\pi_{-\mathcal{A}}$. Interestingly, this strategy has received little attention in the previous literature. Instead, researchers have focused on other  generalized versions of Tit-for-Tat, which cooperate if at least $m$ co-players cooperated in the previous round \cite{boyd:JTB:1988, kurokawa:PRSB:2009}, i.e. $p_{S,j}=0$ if $j<m$ and $p_{S,j}=1$ if $j\ge m$. Unlike ${\it pTFT}$, these threshold strategies neither enforce a linear relation between payoffs, nor do they induce fair outcomes, suggesting that $pTFT$ may be the more natural generalization of Tit-for-Tat for large social dilemmas. \\

\begin{figure}[t]
\centering
\includegraphics[width=15cm]{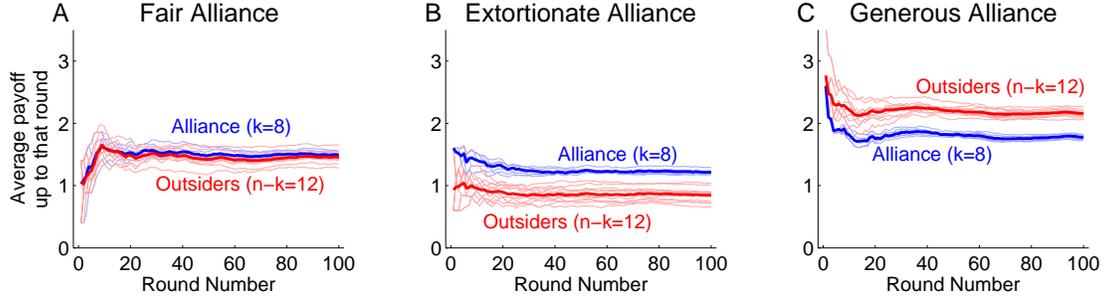}
\caption{Characteristic dynamics of payoffs over the course of the game for three different alliances. Each panel depicts the payoff of each individual group member (thin lines) and the resulting average payoffs (thick lines) for the alliance (blue) and for outsiders (red). (A) An alliance that adopts a fair strategy ensures that the payoff of the allies matches the mean payoff of the outsiders. This does not imply that all outsiders yield the same payoff. (B) For games in which mutual defection leads to the lowest group payoff, extortionate alliances ensure that their payoffs are above average. (C) In games in which mutual cooperation is the social optimum, generous alliances let their co-players gain higher payoffs. The three graphs depict the case of a public goods game with $r=4$, $c=1$, group size $n=20$, and alliance size $k=8$. For the strategies of the outsiders we have used random memory-one strategies, were the cooperation probabilities were independently drawn from a uniform distribution. For the strategies of the allies, we have used (A) $pTFT$, (B) $p^{Ex}$ with $s=0.8$, (C) $p^{Ge}$ with $s=0.8$. } \label{Fig:Illustration2}
\end{figure}

\noindent {\bf Extortionate alliances.} As another interesting subclass of ZD strategies, let us consider strategies that choose the mutual defection payoff as baseline payoff, $l=b_0$, and that enforce a positive slope $0<s_\mathcal{A}<1$. For such strategies, relation [\ref{Eq:PropZD}] becomes $\pi_{-\mathcal{A}}= s_\mathcal{A}\pi_\mathcal{A}+(1-s_\mathcal{A})b_0$, implying that the outsiders only get a fraction $s_\mathcal{A}$ of any surplus over the mutual defection payoff. Moreover, as the slope $s_\mathcal{A}$ is positive, the payoffs $\pi_\mathcal{A}$ and $\pi_{-\mathcal{A}}$ are positively related. As a consequence, the collective best reply for the outsiders is to maximize the allies' payoffs by cooperating in every round. In analogy to Press and Dyson \cite{press:pnas:2012}, we call such alliances {\it extortionate}, and we call the quantity $\chi=1/s_\mathcal{A}$ the extortion factor. Extortionate alliances are particularly powerful in social dilemmas in which mutual defection leads to the lowest group payoff (as in the public goods game and in the volunteer's dilemma): in that case, they enforce the relation $\pi_{-\mathcal{A}} \le \pi_\mathcal{A}$; on average, the allies perform at least as well as the outsiders (as also depicted in Figure~2B). Similarly to the results for fair alliances, extortionate alliances consist of extortionate players: an alliance that enforces the baseline payoff $l=b_0$ and a slope $0<s_\mathcal{A}<1$ requires the allies to use ZD strategies with $l=b_0$ and $0<s<1$, such that each player $i$ individually enforces the relation $\pi_{-i}=s\pi_i+(1-s)b_0$. 

For the specific example of a public goods game, let us consider the ZD strategy $p^{Ex}$ with $l=0$, $\phi=1/c$, and $0<s<1$, for which Eq.~[\ref{Eq:DefZD}] becomes
\begin{equation}
\begin{array}{lcl}
p_{C,j}^{Ex}	&=	&\frac{j}{n-1}-(1-s)\left(\frac{rj}{n}-\frac{n-r}{n}\right)\\[0.3cm]
p_{D,j}^{Ex}	&=	&\frac{j}{n-1}-(1-s)\frac{rj}{n}
\end{array}
\end{equation}
In the limit of $s\rightarrow 1$, these extortionate strategies approach the fair strategy $pTFT$. However, as $s$ decreases from 1, the cooperation probabilities of $p^{Ex}$ are increasingly biased to the own advantage (with the probabilities $p_{D,j}^{Ex}$ decreasing more rapidly than the probabilities $p_{C,j}^{Ex}$). As with fair strategies, such extortionate strategies exist for all repeated social dilemmas. However, in large groups the power of alliances to extort their peers depends on the social dilemma, and on the size of the alliance (as generally described by condition [\ref{Eq:CharPayRelAlliances}]). 

For example, for single-player alliances ($k=1$) in the public goods game, the feasible extortion factors $\chi$ are bounded when groups become large, with $\chi_{\max}=(n-1)r/\big((n-1)r-n\big)$ being the maximum extortion factor (see also \cite{Pan:arxiv:2014}). To be able to enforce arbitrarily high extortion factors, players need to form an alliance such that the fraction of alliance members exceeds a critical threshold. By solving condition ~[\ref{Eq:CharPayRelAlliances}] for the case of extortionate coalitions with infinite extortion factors (i.e., $l=0$ and $s_\mathcal{A}=0$), this critical threshold can be calculated explicitly as
\begin{equation}
\frac{k}{n} \ge \frac{r-1}{r}.
\end{equation}
Only for alliances that have this critical mass, there are no bounds to extortion. \\

\noindent {\bf Generous alliances.} As the benevolent counterpart to extortioners, Stewart and Plotkin were first to describe a set of generous strategies for the iterated prisoner's dilemma \cite{stewart:pnas:2012,stewart:pnas:2013}. Unlike extortioners, generous alliances set the baseline payoff to the mutual cooperation payoff $l=a_{n-1}$, while still enforcing a positive slope $0<s_\mathcal{A}<1$. This results in the payoff relation  $\pi_{-\mathcal{A}} = s_\mathcal{A}\pi_\mathcal{A} + (1-s_\mathcal{A}) a_{n-1}$, such that a generous alliances accept a larger share of any loss (compared to the mutual cooperation payoff $a_{n-1}$). In particular, for games in which mutual cooperation is the optimal outcome (as in the public goods game and in the prisoner's dilemma, but not in the volunteer's dilemma), the payoff of a generous player satisfies $\pi_\mathcal{A}\le \pi_{-\mathcal{A}}$  (see also Fig. 2C depicting the case of a public goods game). As with fair and extortionate alliances, generous alliances consist of players that are individually generous.

For the example of a public goods game, we obtain a particularly simple generous ZD strategy $p^{Ge}$ by setting $l=rc-c$, $\phi=1/c$, and $0<s<1$, such that
\begin{equation}
\begin{array}{lcl}
p_{C,j}^{Ge}	&=	&\frac{j}{n-1}+(1-s)\frac{(n-j-1)r}{n}\\[0.3cm]
p_{D,j}^{Ge}	&=	&\frac{j}{n-1}+(1-s)\frac{(n-j)r-n}{n}
\end{array}
\end{equation}
In parallel to the extortionate strategy discussed before, these generous strategies approach $pTFT$ in the limit of $s=1$, whereas they enforce more generous outcomes for $s<1$.  
Again, generous strategies exist for all social dilemmas, but the extent to which players can be generous depends on the particular social dilemma, and on the size of the alliance.\\ 

\noindent {\bf Equalizers.} As a last interesting class of ZD strategies, let us consider alliances that choose $s=(k-1)/(n-1)$, such that by Eq.~[\ref{Eq:sEff}] the effective slope becomes $s_\mathcal{A}=0$. By Eq.~[\ref{Eq:PropZD}], such alliances enforce the payoff relation $\pi_{-\mathcal{A}}=l$, meaning that they have unilateral control over the mean payoff of the outsiders (for the prisoner's dilemma, such equalizer strategies were first discovered by \cite{boerlijst:AMM:1997}).  However, as with extortionate and generous strategies, equalizer alliances need to reach a critical size to be able to determine the outsiders' payoff; this critical size depends on the particular social dilemma, and on the imposed payoff $l$ (the exact condition can be obtained from [\ref{Eq:CharPayRelAlliances}] by setting $s_\mathcal{A}=0$).

For the example of a public goods game, a single player can only set the co-players' mean score if the group size is below $n\le 2r/(r-1)$. For larger group sizes, players need to form alliances, with 
\begin{equation} \label{Eq:Equalizer}
\frac{k}{n} \ge \frac{(n-2)(r-1)}{n+(n-2)r}
\end{equation} 
being the minimum fraction of alliance members that is needed to dictate the outsiders' payoff. Although the right hand side of Eq.~[\ref{Eq:Equalizer}] is monotonically increasing with group size, equalizer alliances are always feasible; in particular, alliances of size $k=n-1$ can always set the payoff of the remaining player to any value between $0$ and $rc-c$.\\  

\begin{table}[t] {\singlespacing  \small
\begin{tabular}{c|c|ccc}
\Xhline{4\arrayrulewidth}
&&&&\\[-.1cm]
\parbox[c]{1.6cm}{\centering Strategy class} &\parbox[c]{1.4cm}{\centering Typical property} &\parbox[c]{1.2cm}{\centering Prisoner's dilemma}	&\parbox[c]{3cm}{\centering Public goods game}	&\parbox[c]{3cm}{\centering Volunteer's dilemma}\\[0.3cm]
\hline	
&&&&\\[-.1cm]

\parbox[c]{1.6cm}{\centering Fair\\ strategies} &$\pi_{-\mathcal{A}}=\pi_\mathcal{A}$	 &\parbox[c]{1.2cm}{\centering Always exist}	&\parbox[c]{1.2cm}{\centering Always exist}	&\parbox[c]{1.2cm}{\centering Always exist}\\[0.4cm]

\parbox[c]{1.6cm}{\centering Extortionate strategies}       &$\pi_{-\mathcal{A}}\le \pi_\mathcal{A}$	&\parbox[c]{1.2cm}{\centering Always exist}	&\parbox[c]{4.7cm}{\centering In large groups, single players cannot be arbitrarily extortionate, but sufficiently large alliances can be arbitrarily extortionate}	&\parbox[c]{3.9cm}{\centering  Even large alliances cannot be arbitrarily extortionate}\\[0.9cm]

\parbox[c]{1.6cm}{\centering Generous strategies}       &$\pi_{-\mathcal{A}}\ge \pi_\mathcal{A}$ &\parbox[c]{1.2cm}{\centering Always exist}	&\parbox[c]{4.7cm}{\centering In large groups, single players cannot be arbitrarily generous, but sufficiently large alliances can be arbitrarily generous}	&\parbox[c]{3.8cm}{\centering  Do not ensure that own payoff is below average}\\[0.9cm]

Equalizers 	&$\pi_{-\mathcal{A}}=l~$     &\parbox[c]{1.2cm}{\centering Always exist}	&\parbox[c]{4.7cm}{\centering May not be feasible for single players, but is always feasible for sufficiently large alliances}	&\parbox[c]{3.9cm}{\centering Only feasible if the size of the alliance is $k=n-1$, can only enforce $l=b-c$}\\[0.5cm]

\Xhline{4\arrayrulewidth}
\end{tabular}}
\caption{Strategic power of different ZD strategies for three different social dilemmas. In the repeated prisoner's dilemma, single players can exert all strategic behaviors \cite{press:pnas:2012,stewart:pnas:2013,hilbe:plosone:2013b}. Other social dilemmas either require players to form alliances in order to gain sufficient control (as in the public goods game), or they only allow for limited forms of control (as in the volunteer's dilemma).} \label{Tab:SumZD}
\end{table}

\noindent {\bf Strategic power of different ZD strategies.}
Table 1 gives an overview for these four strategy classes for three examples of social dilemmas.  It shows that while generally ZD strategies exist for all group sizes, the power of single players to enforce particular outcomes typically diminishes or disappears in large groups. Forming alliances allows players to increase their strategic scope. The impact of a given alliance, however, depends on the specific social dilemma: while alliances can become arbitrarily powerful in public goods games, their strategic options remain limited in the volunteer's dilemma. 

\begin{figure}[h!]
\centering
\includegraphics[width=7cm]{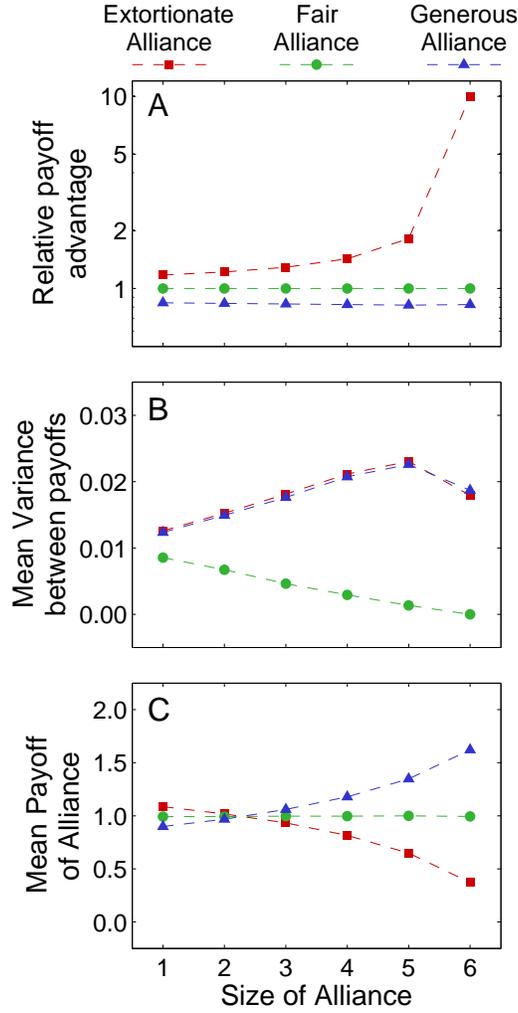}
\caption{The effect of different alliance strategies and various alliance sizes. Each panel shows the outcome of simulated public goods games in which the alliance members interact with $n-k$ random co-players (uniformly taken from the set of memory-one strategies). We compare the success of different alliances along three dimensions: the relative payoff advantage of the alliance (defined as $\pi_\mathcal{A}/\pi_{-\mathcal{A}}$), the payoff inequality within a group (defined as the mean variance between payoffs of all group members), and the absolute payoff of the alliance (as given by $\pi_\mathcal{A}$). Simulations suggest that (A) extortionate alliances gain the highest relative payoff advantages, (B) fair alliances reduce inequality within their group, and (C) sufficiently large generous alliances get the highest payoffs. For the simulations, we have used a public goods game ($r=3$, $c=1$) in a group of size $n=7$; data was obtained by averaging over 500 randomly formed groups. The strategy of the alliance members was $pTFT$, $p^{Ex}$ (with $s=0.85$), and $p^{Ge}$ (with $s=0.85$), respectively.} \label{Fig:DiffZD}
\end{figure}

While fair, extortionate, and generous alliances enforce different payoff relations, simulations suggest that each of these strategy classes has its particular strength when facing unknown opponents (Figure 3). Forming an extortionate alliance gives the allies a relative advantage compared to the outsiders, and by increasing the alliance's size, allies can enforce more extreme relationships. Forming a fair alliance, on the other hand, is an appropriate measure to reduce the payoff inequality within a group -- while the other two behaviors, generosity and extortion, are meant to induce unequal payoffs (to the own advantage, or to the advantage of the outsiders members, respectively), fair players actively avoid generating further inequality by matching the mean payoff of the outsiders.  Generous alliances, however, are most successful in increasing the absolute payoffs. While it is obvious that generous alliances are beneficial for the outsiders (and that this positive effect is increasing in the size of the alliance), Figure 3 suggests that even the allies themselves may benefit from coordinating on a generous alliance strategy. Fair and extortionate alliances are programmed to fight back when being exploited; this is meant to reduce the outsiders' payoffs, but it also reduces the payoffs of the other allies. Therefore, when the alliance has reached a critical size, it is advantageous to agree on a generous alliance strategy instead (but without being overly altruistic), as it helps to avoid self-destructive vendettas. 

This somewhat unexpected strength of generous strategies is in line with previous evolutionary results for the iterated prisoner's dilemma. For this pairwise dilemma, several studies have reported that generosity, and not extortion, is favored by selection \cite{akin:2013,stewart:pnas:2013,hilbe:plosone:2013b}. Such an effect has also been confirmed in a recent behavioral experiment, in which human cooperation rates against generous strategies were twice as high as against extortioners, although full cooperation would have been the humans' best response in both cases \cite{hilbe:natcomm:2014}. Our results suggest that in multiplayer dilemmas, generous alliances are able to induce a similarly beneficial group dynamics. In the Supporting Information we show that if a generous alliance has reached a critical mass, it becomes optimal for outsiders to become generous too (independent of the specific social dilemma, and independent of the strategy of the remaining outsiders). Once this critical mass is achieved, generosity proves self-enforcing.

\section*{Discussion} 

When subjects lack individual power to enforce beneficial outcomes, they can often improve their strategic position by joining forces with others. Herein, we have used and expanded the theory of zero-determinant strategies \cite{press:pnas:2012,akin:2013,stewart:pnas:2013} to explore the role of such alliances in repeated dilemmas. We have found that three key characteristics determine the effect of an alliance of ZD strategists: the underlying social dilemma, the size of the alliance, and the strategy of the allies. While subjects typically have little influence to transform the underlying dilemma, we have shown that they can considerably raise their strategic power by forming larger alliances, and they can achieve various objectives by choosing appropriate strategies. 

Our approach is based on the distinction between alliance members (who agree on a joint ZD strategy), and outsiders (who are not restricted to any particular strategy, and who may form an alliance themselves). This distinction allowed us to show the existence of particularly powerful alliances, and to discuss their relative strengths. As an interesting next step of research, we plan to investigate how such alliances are formed in the first place (which is typically at the core of traditional models of coalitions, e.g. \cite{peleg:book:2003}), and whether evolutionary forces would favor particular alliances over others \cite{Mesterton-Gibbons:jtb:2011}. The results presented herein suggest that subjects may have various motives to join forces. As particular examples, we have highlighted extortionate alliances (who aim for a relative payoff advantage over the outsiders), fair alliances (who aim to reduce the inequality within their group), and generous alliances (who are able to induce higher payoffs as they avoid costly vendettas after accidental defections). Whether such alliances emerge and whether they are stable thus needs to be addressed in light of the respective aim of the alliance: when subjects are primarily interested in low inequality, then forming a fair alliance is an effective means to reach this aim; and once a fair alliance is formed, inequity-averse subjects have little incentives to leave (even if leaving the alliance would allow them to gain higher payoffs). 

While we have focused on the effects of alliances in multiplayer social dilemmas, it should be noted that our results on ZD strategies also apply for solitary alliances, consisting of single players only. 
Thus, even if players are unable to coordinate on joint strategies, zero-determinant strategies are surprisingly powerful. They allow players to dictate linear payoff relations, irrespective of the specific social dilemma being played, irrespective of the group size, and irrespective of the counter-measures taken by the outsiders. In particular, we have shown that any social dilemma allows players to be fair, extortionate, or generous. At the same time, zero-determinant strategies are remarkably simple. For example, in order to be fair in a public goods game (or in a volunteer's dilemma), players only need to apply a rule called proportional Tit-for Tat $pTFT$: if $j$ of the $n-1$ other group members cooperated in the previous round, then cooperate with probability $j/(n-1)$ in the following round. Extortionate and generous strategies can be obtained in a similar way, by slightly modifying $pTFT$ to the own advantage or to the advantage of the outsiders. 

While these results were derived for the special case of infinitely repeated games, they can be extended to the more realistic finite case. In finitely repeated games, end-game effects may prevent alliances to enforce a perfect linear relation between payoffs;  but it is still possible to enforce an arbitrarily strong correlation between payoffs, provided that the game is repeated sufficiently often. Similarly, we show in the Supporting Information, that it is not necessary that all alliance members coordinate on the same ZD strategy, and that different alliance members may apply different strategies. However, we have focused here on the case of symmetric alliances with a joint strategy, because they are most powerful: any payoff relationship that can be enforced by asymmetric alliances with different ZD strategies, can also be enforced by a symmetric alliance.

Overall, our results reveal how single players in multiplayer games can increase their strategic power by forming beneficial alliances with others, helping them to regain control in large-scale social dilemmas.

\subsection*{Acknowledgments}
CH gratefully acknowledges generous funding by the Schr\"odinger stipend J3475 of the Austrian Science Fund.

{\small
\setlength{\bibsep}{0\baselineskip}

\newpage
\renewcommand{\theequation}{S\arabic{equation}}
\setcounter{equation}{0}

{\centering \sffamily\textbf{\Large Supporting Information: Zero-determinant alliances in multiplayer social dilemmas\\}}
\author{\parbox[c]{14.5cm}{\center{\normalsize Christian Hilbe$^1$, Arne Traulsen$^2$, Bin Wu$^2$ \& Martin A. Nowak$^{1,3}$}\\[0.2cm]
\normalsize $^1$ Program for Evolutionary Dynamics, Harvard University, Cambridge, MA~02138, USA\\[0.1cm]
$^2$ EvolutionaryTheory Group, Max-Planck-Institute for Evolutionary Biology, August-Thienemann-Stra{\ss}e 2, 24306 Pl\"{o}n, Germany\\[0.1cm]
$^3$ {Department of Organismic and Evolutionary Biology, Department of Mathematics, Harvard University, Cambridge, MA 02138}}}

~\\[1cm]
\noindent In the following, we develop a theory of zero-determinant strategies (ZD strategies) and alliances for general multiplayer social dilemmas. We begin by defining the setup of our model of repeated social dilemmas (Section 1). Thereafter, we derive the existence and the properties of ZD strategies for solitary alliances with a single player (Section 2), and then for general alliances with an arbitrary number of players (Section 3). Moreover, we explore which ZD strategies give rise to stable Nash equilibria, and we discuss which alliances are self-enforcing when subjects strive for high payoffs (Section 4). As applications, we study alliances in the repeated public goods game and in the repeated volunteer's dilemma (Section 5). The appendix contains the proofs for our results.

\tableofcontents
\newpage

\section{Setup of the model: Repeated multiplayer dilemmas}

We consider repeated social dilemmas between $n$ players (as illustrated in Fig.~1 of the main text). In each round, players may either cooperate (C) or defect (D), and the players' payoffs for each round depend on their own action, and on the number of cooperators among the other group members. Specifically, in a group with $j$ other cooperators, a cooperator receives the payoff $a_j$, whereas a defector obtains $b_j$. To qualify as a social dilemma, we assume that one-shot payoffs satisfy the following three conditions: 
\begin{description}
\item{({\it i})} Independent of the own action, players prefer their co-players to be cooperative,
\begin{equation}
a_{j+1}\ge a_j ~~~\text{and} ~~~ b_{j+1}\ge b_j ~~\text{for all $j$ with $0\le j <n-1$.}
\end{equation}
\item{({\it ii})} Within each mixed group, defectors strictly outperform cooperators,
\begin{equation}
b_{j+1} > a_j ~~\text{for all $j$ with $0\le j <n-2$.}
\end{equation}
\item{({\it ii})} Mutual cooperation is preferred over mutal defection, 
\begin{equation}
a_{n-1}>b_0.
\end{equation}
\end{description}
As particular examples of such social dilemmas, we discuss the linear public goods game (e.g. \cite{ledyard:bookchapter:1995}) and the volunteer's dilemma \cite{diekmann:jcr:1985} in Section~5. 

Let us assume that the social dilemma is repeated for $M$ rounds (in the main manuscript, we have focused on the special case $M\rightarrow \infty$). For such repeated games, a player's strategy needs to specify how to act in given round, depending on the outcomes of the previous rounds. Given the strategies of all group members, let us denote player $i$'s expected payoff in round $m$ as $\pi_i(m)$, and the average payoff of his co-players as $\pi_{-i}(m)=\sum_{j\neq i} \pi_j(m)/(n-1)$. If the social dilemma is repeated for  $M$ rounds, we define payoffs for the repeated game as the average payoff per round,
\begin{equation} 
\begin{array}{lll} \label{Eq:payoffsSI}
\pi_i	&= &\displaystyle \frac{1}{M} \sum_{m=0}^{M-1} \pi_i(m).\\[0.5cm]
\pi_{-i}	&= &\displaystyle \frac{1}{M} \sum_{m=0}^{M-1} \pi_{-i}(m).
\end{array}
\end{equation}
For infinitely repeated games, payoffs are then defined by taking the limit $M\rightarrow \infty$ For simplicity, we assume that these limits exist (which holds, for example, under the realistic assumption that players sometimes commit implementation errors, \cite{sigmund:book:2010}).

While in general, strategies for repeated games can be arbitrarily complicated, ZD strategies are a subset of a particularly simple class of strategies called memory-one strategies \cite{nowak:nature:1993,sigmund:book:2010,press:pnas:2012}. In finitely repeated $n$-player games, such strategies can be written as a vector
\begin{equation}
\tilde{\mathbf{p}}=\big( p_0;~ p_{C,n-1},p_{C,n-2},\ldots,p_{C,0};~p_{D,n-1},p_{D,n-2}\ldots,p_{D,0}\big),
\end{equation}
where $p_0$ is the player's probability to cooperate in the first round, and $p_{S,j}$ is the probability to cooperate in round $m\ge2$, given that the player previously played $S\in\{C,D\}$, and that $j$ of his co-players cooperated. In infinitely repeated games, the outcome of the first round can often be neglected; in that case, one may drop the entry for $p_0$ from the representation of memory-one strategies \cite{press:pnas:2012,stewart:pnas:2012,hilbe:pnas:2013,akin:2013,stewart:pnas:2013,hilbe:plosone:2013b,szolnoki:pre:2014,Pan:arxiv:2014}, yielding 
\begin{equation} \label{eq:mem1SI}
\mathbf{p}=\big(p_{C,n-1},p_{C,n-2},\ldots,p_{C,0};~p_{D,n-1},p_{D,n-2}\ldots,p_{D,0}\big).
\end{equation}
If the group contains only players with memory-strategies, the calculation of payoffs according to [\ref{Eq:payoffsSI}] becomes particularly simple. In that case, the repeated game can be described as a Markov chain, as the outcome in the next round only depends on the outcome of the previous round \cite{nowak:nature:1993,hauert:prsb:1997,sigmund:book:2010}. While the assumption of memory-one strategies often simplifies calculations, we will show below that the properties of ZD strategies hold irrespective of the strategies of the other group members (in particular, ZD strategists do not require their co-players to apply memory-one strategies).

\section{ZD strategies for solitary players}
As specified in the main text, ZD strategies are particular memory-one strategies. We say that a player adopts a ZD strategy, if the player chooses three constants $l$, $s$, and $\phi\neq 0$, and then calculates the respective cooperation probabilities in [\ref{eq:mem1SI}] as
\begin{equation} \label{eq:defzdSI}
\begin{array}{llc}
p_{C,j}	&=	&\displaystyle 1+\phi \Big[(1-s)(l-a_j)-\frac{n-j-1}{n-1}(b_{j+1}-a_j)\Big]\\[0.25cm]
p_{D,j} 	&=	&\displaystyle ~\phi \Big[(1-s)(l-b_j)+\frac{j}{n-1}(b_j-a_{j-1})\Big].
\end{array}
\end{equation}
It is the following property of ZD strategies that is central to our analysis: 

\begin{Prop}[ZD strategies enforce a linear relation between payoffs.] \label{Prop:zdSI}
Suppose player $i$ applies a ZD strategy with parameters $l,s$, and $\phi> 0$ in a repeated game with $M$ rounds. Then payoffs obey the relation 
\begin{equation} \label{Eq:propZDSI1}
\big| \pi_{-i}-s\pi_i-(1-s)l \big| \le \frac{1}{\phi M}.
\end{equation}
In particular, in the case of infinitely repeated games, $M\rightarrow \infty$, payoffs satisfy
\begin{equation} \label{Eq:propZDSI2}
\pi_{-i}=s\pi_i+(1-s)l.
\end{equation}
\end{Prop}

\begin{figure}[tb] 
\centering
\includegraphics[height=6.9cm]{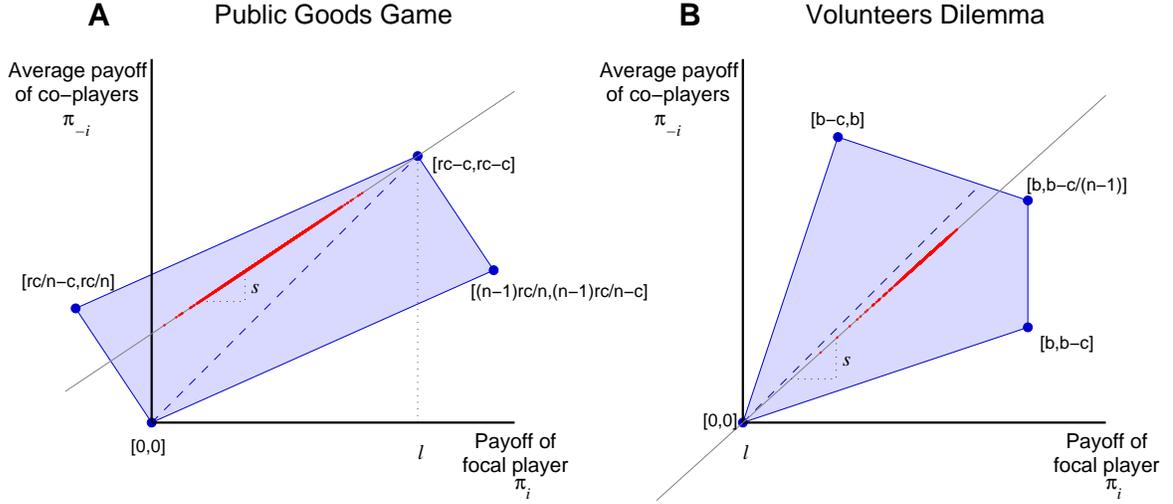}~\\[-.5cm]
\caption{Illustration of ZD-strategies for  (A) the linear public goods game and (B) the volunteer's dilemma. The blue-shaded area represents all feasible payoffs, with the $x$-axis representing the payoff of player $i$, and with the $y$-axis representing the mean payoff of $i$'s co-players. The dashed diagonal gives the payoff combinations for which $\pi_i=\pi_{-i}$. In both graphs, the strategy of player~$i$ is fixed to some zero-determinant strategy, whereas for the co-players we have sampled $10^4$ random memory-one strategies. Red dots represent the resulting payoff combinations, and the grey line gives the prediction according to Eq.~\eqref{Eq:propZDSI2}. For both graphs, we have considered an infinitely repeated game in a group of size $n=4$. Parameters: (A) Public goods game with $r=2.4$ and $c=1$. For the strategy of player $i$ we have used a generous ZD strategy with parameters $l=rc-c$, $s=2/3$, $\phi=1/2$. (B) Volunteer's dilemma with $b=1.5$, $c=1$; player $i$ applies an extortionate strategy with parameters $l=0$, $s=9/10$, $\phi=1/2$.}\label{Fig:S1}
\end{figure}

\noindent It is worth to highlight several features of Proposition \ref{Prop:zdSI}. First, the Proposition makes no assumptions on the strategies of the other group members, nor does it make restrictions on the specific social dilemma. Player $i$ can thus enforce linear relations of the form [\ref{Eq:propZDSI1}] and [\ref{Eq:propZDSI2}], independent of the behavior of $i$'s co-players, and independent of the exact strategic situation (in fact, the proof of Proposition 1 does not make use of the assumption that the game is a social dilemma). In Figure \ref{Fig:S1}, we illustrate this result for two different social dilemmas (the public goods game and the volunteer's dilemma), and for two different ZD strategies (a generous and an extortionate ZD strategy). Second, while we have focused on infinitely repeated games in the main text, relation [\ref{Eq:propZDSI1}] shows how our results generalize for finitely many rounds. Even in the finite case, player $i$ can enforce an arbitrarily strong correlation between the players' payoffs, provided that the round number $M$ is sufficiently large compared to the parameter $\phi$. Third, Proposition \ref{Prop:zdSI} gives a natural interpretation for the three parameters $l$, $s$ and $\phi$. The baseline payoff $l$ corresponds to the players' payoffs if all group members apply the same ZD strategy (in which case $\pi_{-i}=\pi_i$); the slope $s$ determines how strong the payoffs $\pi_i$ and $\pi_{-i}$ are  related; and by [\ref{Eq:propZDSI1}] the parameter $\phi$ determines the rate of convergence with $M$. 

A player cannot enforce arbitrary payoff relations [\ref{Eq:propZDSI2}] because the parameters $l$, $s$ and $\phi$ need to be set such that all cooperation probabilities in [\ref{eq:defzdSI}] are in the unit interval. We thus say that a payoff relation ($l$,$s$) is enforceable, if there is a $\phi>0$ such that $0\le p_{S,j}\le 1$ for all $S\in\{C,D\}$ and all $j$ with $0\le j\le n-1$. The following gives a characterization of the possible payoff relations that a single player can enforce. 

\begin{Prop}[Characterization of enforceable payoff relations] \label{Prop:charpayrelSI}
For a given social dilemma, the payoff relation ($l,s$) is enforceable if and only if either $s=1$ or
\begin{equation} \label{eq:charrelSI}
\small
\displaystyle
\begin{array}{ccccc}
{\displaystyle \!\! \max_{0\le j\le n-1}} \left\{b_j-\frac{j}{(1-s)(n-1)}(b_j-a_{j-1})\right\} \!\!  &\!\!  \le \!\! 	&\!\! l \!\! 	&\!\!  \le \!\!  &\!\! {\displaystyle \min_{0\le j\le n-1}} \!\!  \left\{a_j+\frac{n-j-1}{(1-s)(n-1)}(b_{j+1}-a_j)\right\}.
\end{array}
\end{equation}
In particular, enforceable payoff relations satisfy $-\frac{1}{n-1}\le s \le 1$, and if $s<1$ then $b_0\le l\le a_{n-1}$. 
\end{Prop}

\begin{figure}[tb] 
\centering
\includegraphics[height=6.9cm]{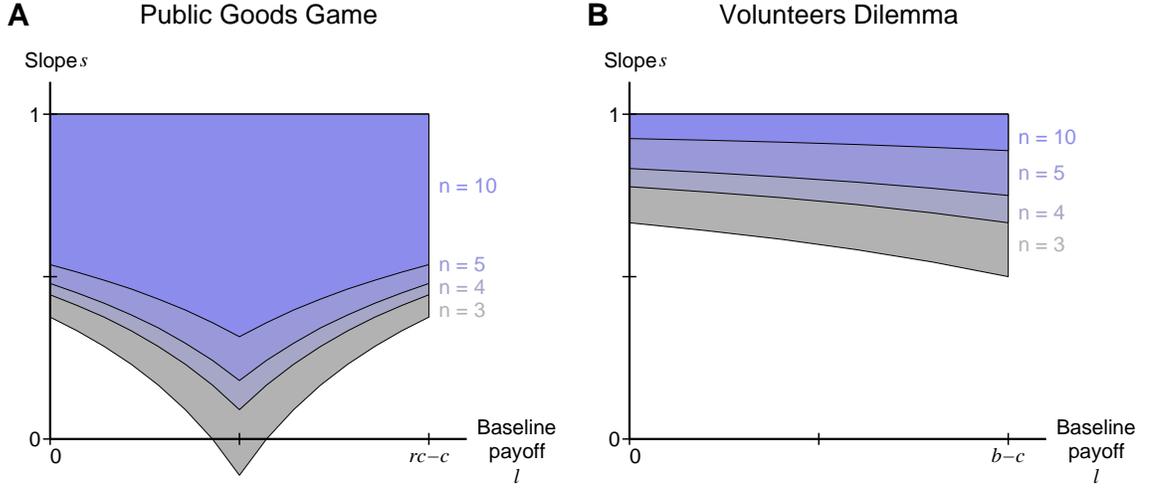}
\caption{Enforceable payoff relations for (A) the linear public goods game and (B) the volunteer's dilemma. A pair ($l,s$) is enforceable for a given group size $n$ if the point is within the respectively shaded area. The set of enforceable pairs for large $n$ is a subset of the the respective set for smaller $n$, i.e. the set of enforceable pairs shrinks with increasing group size. Parameters: (A) Linear public goods game with $r=2.4$, $c=1$. (B) Volunteer's dilemma with $b=1.5$, $c=1$.}\label{Fig:spaceZD}
\end{figure}

\noindent As a corollary, we note that if for a given slope $s$ the baseline payoffs $l_1$ and $l_2$ are enforceable, then so is any baseline payoff between $l_1$ and $l_2$.
Figure~\ref{Fig:spaceZD} gives an illustration of the enforceable payoff relations, again for the two examples of a public goods game and a volunteer's dilemma. In particular, it shows that the space of enforceable payoff relations shrinks with group size. Proposition \ref{Prop:charpayrelSI} confirms that this effect can be observed for all social dilemmas (more specifically, for all games that satisfy $b_{j+1}>a_j$). Thus, in larger groups, a single player has less strategic options to enforce particular payoff combinations.

\section{ZD strategies for alliances} 

To extend this result for single subjects to alliances, let us suppose the first $k$ players form an alliance $\mathcal{A}=\{1,\ldots, k\}$, and they agree on a joint zero-determinant with parameters $l,s,\phi$. Thus, each of the allies $i\in \mathcal{A}$ enforces the relation $\pi_{-i}=s\pi_i+(1-s)l$. Summing up over all allies $1\le i \le k$ yields
\begin{equation}
\displaystyle \frac{k-1}{n-1}(\pi_1+\ldots+\pi_k)+\frac{k}{n-1}(\pi_{k+1}+\ldots+\pi_n)= s(\pi_1+\ldots+\pi_k)+k(1-s)l.
\end{equation}
Rearranging these terms then implies
\begin{equation} \small
\frac{\pi_{k+1}+\ldots+\pi_n}{n-k}= \frac{s(n-1)-(k-1)}{n-k}\left(\frac{\pi_1+\ldots+\pi_k}{k}\right)+\left(1- \frac{s(n-1)-(k-1)}{n-k}\right)l,
\end{equation}
or, in a more intuitive notation,
\begin{equation} \label{eq:zdallSI}
\pi_{-\mathcal{A}}=s_\mathcal{A}\pi_\mathcal{A}+(1-s_\mathcal{A})l,
\end{equation}
with $\pi_\mathcal{A}$ being the mean payoff of the allies, $\pi_{-\mathcal{A}}$ being the mean payoff of the outsiders, and 
\begin{equation} \label{eq:saSI}
s_\mathcal{A}=\frac{s(n-1)-(k-1)}{n-k}
\end{equation}
being the effective slope of the alliance. Thus the results for a single ZD strategist naturally extend to the case of arbitrarily many allies. For the set of enforceable payoff relations ($l,s_\mathcal{A}$) we can give an analogous characterization:

\begin{Prop}[Characterization of enforceable payoff relations for alliances] \label{prop:allianceSI}
For a given social dilemma, an alliance with $k$ members can enforce the payoff relation ($l,s_\mathcal{A}$) if and only if either $s_\mathcal{A}=1$ or
\begin{equation} \label{eq:charallSI} \small
\begin{array}{ccccc}
{\displaystyle \!\! \max_{0\le j\le n-1}} \!\left\{b_j-\frac{j}{(1-s_\mathcal{A})(n-k)}(b_j-a_{j-1})\right\} \!\!  &\!\!  \le \!\!\! 	&\!\! l \!\! 	&\!\!\!  \le \!\!  &\!\! {\displaystyle \min_{0\le j\le n-1}} \!\!  \left\{a_j+\frac{n-j-1}{(1-s_\mathcal{A})(n-k)}(b_{j+1}-a_j)\right\}.\\
\end{array}
\end{equation}
In particular, if a given payoff relation ($l,s_\mathcal{A}$) can be enforced by some alliance of ZD strategists (possibly by using different ZD strategies), then it can be enforced by a symmetric alliance (where all players apply the same ZD strategy). Moreover, enforceable payoff relations satisfy $-\frac{1}{n-k}\le s_\mathcal{A} \le 1$, and if $s_\mathcal{A}<1$ then $b_0\le l\le a_{n-1}$. 
\end{Prop}

\noindent As a consequence of Proposition \ref{prop:allianceSI} we note that space of enforceable payoff relations ($l$,$s_\mathcal{A}$) increases with the size of the alliance $k$ (this holds true for any game in which $b_{j+1}>a_j$ for all~$j$). Thus, while Proposition \ref{Prop:charpayrelSI} has led to the conclusion that players lose their strategic options as groups become large, Proposition \ref{prop:allianceSI} suggests that players can offset this loss of strategic power in large groups by forming alliances.

\section{Self-enforcing alliances}

Let us next explore which ZD-strategies form a Nash equilibrium (i.e., which ZD strategies can be stably maintained in a population, assuming that players aim at high payoffs). Having the previous section in mind, we may ask equivalently: suppose there is an alliance with $k=n-1$ players, applying some ZD-strategy. In which case would it be optimal for the remaining player to apply this ZD-strategy too? To respond to this question, let us consider an alliance $\mathcal{A}=\{1,\ldots,n-1\}$ applying a ZD strategy with parameters $l,s,\phi$.  As a minimum requirement for this strategy to be an equilibrium, the remaining player $n$ must not have an incentive to choose a different ZD-strategy with parameters $l_n,s_n,\phi_n$. By [\ref{Eq:payoffsSI}], such a deviating player enforces the payoff relation
\begin{equation}
\pi_\mathcal{A}=s_n\pi_n+(1-s_n)l_n,
\end{equation}
whereas, by [\ref{eq:zdallSI}], the alliance $\mathcal{A}$ enforces the payoff relation
\begin{equation}
\pi_n=s_\mathcal{A}\pi_\mathcal{A}+(1-s_\mathcal{A})l.
\end{equation}
Assuming $s_n<1$, this system of linear equations yields the following payoff for the deviating player
\begin{equation}
\pi_n=\frac{l(1-s_\mathcal{A})+l_ns_\mathcal{A}(1-s_n)}{1-s_\mathcal{A}s_n}.
\end{equation}
By adhering to the alliance strategy, player $n$ would obtain the payoff $l$. Thus, deviating from the alliance strategy yields an advantage if
\begin{equation}
\pi_n-l=\frac{(l_n-l)s_\mathcal{A}(1-s_n)}{1-s_\mathcal{A}s_n}>0.
\end{equation}
We can distinguish three cases:
\begin{enumerate}
\item $s_\mathcal{A}=0$: In that case $\pi_n-l=0$, i.e. player $n$ cannot improve his payoff by deviating unilaterally.
\item $s_\mathcal{A}>0$: In that case $\pi_n-l>0$ if and only if $l_n>l$. Thus, to be stable against invasion of other ZD strategists, the alliance needs to apply a strategy with the maximum possible baseline payoff, $l=a_{n-1}$.
\item $s_\mathcal{A}<0$: then $\pi_n-l>0$ if and only if $l_n-l<0$. Thus the ZD-strategy of the alliance is stable against invasion by other ZD-strategists if and only if $l$ is minimal, i.e. $l=b_0$. 
\end{enumerate}
As the following Proposition shows, this result also holds if deviating players are not restricted to zero-determinant strategies.

\begin{Prop}[Pure Nash equilibria among ZD-strategies] \label{Prop:nashSI}
Consider a ZD-strategy with parameters $l$, $s$, $\phi$, and a social dilemma in which mutual cooperation is the best outcome for the group, whereas mutual defection is the worst possible outcome, 
\begin{equation}
b_0\le \min_{0\le j\le n} \frac{ja_{j-1}+(n-j)b_{j}}{n} \le a_{n-1}.
\end{equation}
Let $s_\mathcal{A}$ be defined as in Eq. [\ref{eq:saSI}], for $k=n-1$. Then the ZD-strategy is a Nash equilibrium if and only if one of the following three cases holds:
\begin{enumerate}
\item $s_\mathcal{A}=0$, i.e., if applied by $n-1$ players, the ZD-strategy acts as an equalizer.
\item $s_\mathcal{A}>0$ and $l=a_{n-1}$, i.e., if applied by $n-1$ players, the ZD-strategy is generous. 
\item $s_\mathcal{A}<0$ and $l=b_0$, i.e., if applied by $n-1$ players the ZD-strategy is selfish. 
\end{enumerate}
\end{Prop}

\noindent Some observations are in order. First, the three conditions in Propostion \ref{Prop:nashSI} do not depend on $\phi$. Whether a ZD-strategy is stable depends on the enforced payoff relation only (i.e., on $l$ and $s$), but not on the exact strategy that gives rise to this payoff relation. 

Second, the stability of a ZD-strategy does not depend  on the sign of $s$, but on the sign of $s_\mathcal{A}=(n-1)s-(n-2)$. In particular, a generous ZD-strategy with parameters $l=a_{n-1}$ and $s>0$ is only stable if it is not too generous, $s>(n-2)/(n-1)$. Thus, only for pairwise dilemmas (with $n=2$), {\it any} generous ZD strategy is stable \cite{akin:2013,stewart:pnas:2013}; as the group size increases, generous strategies need to approach the fair strategies (with $s=1$) in order to be stable. 

Third, Proposition \ref{Prop:nashSI} also allows us to respond to the question when an alliance is self-enforcing (in the sense that outsiders prefer to join the alliance, once the alliance has exceeded a critical size~$k$). If players are only interested in high absolute payoffs (and not in fairness, or in relative payoff advantages), then alliances either need to be equalizers, generous, or selfish in order to be self-enforcing.

\section{Applications}

\subsection{Public goods games}
As one particular example of a multiplayer dilemmas, let us first study the properties of repeated public goods games. In a public goods game, each player of a group can cooperate by contributing an amount $c>0$ to a public pool. Total contributions are multiplied by a factor $r$ with $1<r<n$, and evenly shared among all group members. Thus, payoffs are given by
\begin{equation}
a_j=\frac{(j+1)rc}{n}-c,~~~\text{and}~~~b_j=\frac{jrc}{n}.
\end{equation}
For solitary alliances, some of the properties of ZD strategies for public goods games have been recently described by \cite{Pan:arxiv:2014}, using an independent approach. Here we complement and extend these results; in particular, we discuss the effect of alliance sizes $k>1$. 

Because the payoffs of the public goods game are linear in the number of players $j$, the characterization of enforceable payoff relations according to Eq. [\ref{eq:charallSI}] becomes particularly simple (as only the boundary cases $j=0$ and $j=n-1$ need to be considered). We conclude that alliances of size $k$ can enforce a linear relation with parameters ($l$,$s_\mathcal{A}$) if either $s_\mathcal{A}=1$ or
\begin{equation} \label{eq:charpggSI} \small
\begin{array}{ccccc}
\displaystyle \max \left\{0,\frac{(n-1)rc}{n}-\frac{(n-1)c}{(1-s_\mathcal{A})(n-k)}\right\} \!\!\!\!  &\!\!  \le \!\!\! \!	&\!\! l \!\! 	&\!\!\!\!  \le \!\!  &\displaystyle \!\!\! \min  \left\{rc\!-\!c,\frac{(r-n)c}{n}+\frac{(n-1)c}{(1-s_\mathcal{A})(n-k)}    \right\}.\\
\end{array}
\end{equation}

\subsubsection{Solitary players} 
For $k=1$, the condition [\ref{eq:charpggSI}] simplifies even further; a single player can enforce a linear relation with parameters ($l,s$) if and only if either $s=1$, or the following two inequalities are satisfied
\begin{equation} \label{eq:charpgg1SI}
\begin{array}{ccccc}
0	&\le &l	&\le	&rc-c\\[0.3cm]
\displaystyle \frac{(n-1)rc}{n}-\frac{c}{1-s}	&\le	&l	&\le	&\displaystyle \frac{(r-n)c}{n}+\frac{c}{1-s}
\end{array}
\end{equation}
Figure \ref{Fig:spaceZD} shows the set of all pairs ($l,s$) that satisfy the above constraints for various group sizes~$n$. We get the following conclusions for the existence of extortionate strategies, generous strategies, and equalizers:

\begin{enumerate}
\item{{\bf Extortionate strategies} ($l=b_0=0$)}. By the inequalities [\ref{eq:charpgg1SI}], extortionate strategies need to satisfy
\begin{equation} \label{eq:extpggSI}
s\ge \frac{(n-1)r-n}{(n-1)r}.
\end{equation}
In particular, for $r> n/(n-1)$, solitary players cannot enforce arbitrarily high extortion factors $\chi=1/s$. Instead the set of feasible extortion factors is bounded from above by 
\begin{equation}
\chi_{\max}=\frac{(n-1)r}{(n-1)r-n}>1.
\end{equation}
For large group sizes, $n\rightarrow \infty$, this implies a maximum extortion factor $\chi_{\max}=r/(r-1)$.
\item{{\bf Generous strategies} ($l=a_{n-1}=rc-c$).} By the inequalities [\ref{eq:charpgg1SI}], the slope of a generous ZD strategies needs to satisfy the same constraint [\ref{eq:extpggSI}] as extortionate strategies. 
\item {{\bf Equalizers} ($s=0$).} For equalizers, the inequalities [\ref{eq:charpgg1SI}] imply there are three regimes: \\[-.8cm]
\begin{enumerate}
\item If $r\le n/(n-1)$, all baseline payoffs $0\le l\le rc-c$ can be enforced.
\item If $n/(n-1)<r\le n/(n-2)$, only a limited subset of baseline payoffs $0<l<rc-c$ can be enforced.
\item If $r>n/(n-2)$, there are no equalizers.\\[-.8cm] 
\end{enumerate}
In particular, we conclude that for a given multiplication factor $r>1$ the set of equalizer strategies disappears as groups become large.
\end{enumerate}

\subsubsection{Alliances} 
By [\ref{eq:charpggSI}], alliances with $k>1$ can enforce a linear relation with parameters ($l,s_\mathcal{A}$) if and only if either $s_\mathcal{A}=1$, or if the two following inequalities hold:
\begin{equation} \label{eq:charpgg2SI}
\begin{array}{ccccc}
0	&\le &l	&\le	&rc-c\\[0.3cm]
\displaystyle \frac{(n-1)rc}{n}-\frac{(n-1)c}{(1-s_\mathcal{A})(n-k)}	&\le	&l	&\le	&\displaystyle \frac{(r-n)c}{n}+\frac{(n-1)c}{(1-s_\mathcal{A})(n-k)}.
\end{array}
\end{equation}
For the special cases of extortionate strategies, generous strategies, and equalizers, these inequalities allow us to derive the following conclusions:
\begin{enumerate}
\item{{\bf Extortionate strategies} ($l=b_0=0$)}. The inequalities [\ref{eq:charpgg2SI}] can be rewritten as a critical threshold on the fraction of alliance members that is needed to enforce a certain slope $s_\mathcal{A}$, 
\begin{equation} \label{eq:extpgg2SI}
\frac{k}{n}\ge \frac{r(1-s_\mathcal{A})-1}{r(1-s_\mathcal{A})}.
\end{equation}
In particular, if an alliance wants to enforce arbitrarily high extortion factors $\chi\rightarrow \infty$, then $s_\mathcal{A}=1/\chi \rightarrow 0$ and the critical threshold becomes $k/n\ge (r-1)/r$.
\item{{\bf Generous strategies} ($l=a_{n-1}=rc-c$).} The inequalities [\ref{eq:charpgg2SI}]  lead to the same threshold for $k/n$ as in the case of extortionate strategies, as given in [\ref{eq:extpgg2SI}].
\item {{\bf Equalizers} ($s=0$).} For equalizers, the inequalities [\ref{eq:charpgg2SI}]  lead to two critical thresholds; in order to be able to set the payoffs of the outsiders to {\it any} value between $0\le l \le rc-c$, the fraction of the allies needs to satisfy
\begin{equation}
\frac{k}{n}\ge \frac{r-1}{r}.
\end{equation}
However, to be able to set the payoffs of the outsiders to {\it some} value between $0\le l \le rc-c$, the number of allies only needs to exceed
\begin{equation}
\frac{k}{n}\ge \frac{(n-2)(r-1)}{n+(n-2)r}.
\end{equation}
\end{enumerate}

\subsection{Volunteer's Dilemma}
In the volunteer's dilemma, at least one of the players needs to cooperate and pay a cost $c>0$, in order for all group members to derive a benefit $b>c>0$. Thus, the payoffs are given by
\begin{equation}
a_j=b-c~~\text{for all}~j,~~\text{and}~~b_j=b~~\text{if}~~j\ge 1~~\text{and}~~b_0=0.
\end{equation}
To characterize the enforceable payoff relations, we note that by condition [\ref{eq:charallSI}] exactly those parameters $l$ and $s_\mathcal{A}$ can be enforced for which
either $s_\mathcal{A}=1$ or
\begin{equation} \label{eq:charvdSI} \small
\begin{array}{ccccc}
\displaystyle \max \left\{0,b-\frac{c}{(1-s_\mathcal{A})(n-k)}\right\} \!\!\!\!  &\!\!  \le \!\!\! \!	&\!\! l \!\! 	&\!\!\!\!  \le \!\!  &\displaystyle \!\!\! b-c.\\
\end{array}
\end{equation}
In the special case of extortionate strategies ($l=0$), this implies that a given slope $s_\mathcal{A}$ can only be enforced by an alliance of ZD strategists if 
\begin{equation}
\frac{k}{n}\ge 1-\frac{1}{n}\frac{c}{(1-s_\mathcal{A})b}.
\end{equation}
In particular, it follows that even large alliances cannot be arbitrarily extortionate (setting $s_\mathcal{A}=0$ on the right hand side implies that such alliances would need to satisfy $k>n-1$).
Instead, the maximum extortion factor for an alliance of ZD strategists is given by $\chi_{\max}=b/(b-c)$. 

For equalizers ($s=0$),  the inequalities in [\ref{eq:charvdSI}] imply that alliances can only unilaterally determine the outsiders' payoffs if $k=n-1$, in which case only the baseline payoff $l=b-c$ is enforceable.

\appendix
\section{Appendix: Proofs}
\begin{proof}[\bf Proof of Proposition \ref{Prop:zdSI}]
The proof is an extension of previous proofs for pairwise games \cite{hilbe:2013,hilbe:natcomm:2014} to the case of multiplayer games. It is useful to introduce some notation. Let $g^i_{S,j}$ be $i$'s payoff in a given round if player $i$ chooses action $S\in\{C,D\}$, and if $j$ of the co-players cooperate, that is
\begin{equation}
g^i_{S,j}=\left\{
\begin{array}{ll}
a_j	&\text{if~} S=C\\
b_j	&\text{if~} S=D.
\end{array}
\right.
\end{equation}
Similarly, let $g^{-i}_{S,j}$ be the average payoff of the other group members in that case, that is
\begin{equation}
g^{-i}_{S,j}=\left\{
\begin{array}{ll}
\displaystyle \frac{ja_j+(n-j-1)b_{j+1}}{n-1}	&\text{if~} S=C\\[0.4cm]
\displaystyle \frac{ja_{j-1}+(n-j-1)b_j}{n-1}	&\text{if~} S=D.
\end{array}
\right.
\end{equation}
Moreover, let $v_{S,j}(m)$ be the probability that in round $m$, the focal player $i$ chooses $S$ and that $j$ of his co-players cooperate. Let us collect these numbers and write them as vectors,
\begin{equation}
\begin{array}{ccl}
\mathbf{g_i}	&=	&\big(g_{C,n-1}^i,\ldots,g_{C,0}^i,g_{D,n-1}^i,\ldots,g_{D,0}^i\big)\\[0.3cm]
\mathbf{g_{-i}}	&=	&\big(g_{C,n-1}^{-i},\ldots,g_{C,0}^{-i},g_{D,n-1}^{-i},\ldots,g_{D,0}^{-i}\big)\\[0.3cm]
\mathbf{v}(m)	&=	&\big(v_{C,n-1}(m),\ldots,v_{C,0}(m),v_{D,n-1}(m),\ldots,v_{D,0}(m)\big)^T.
\end{array}
\end{equation}
With this notation, player $i$'s expected payoff in round $m$ can be written as $\pi_i(m)=\mathbf{g_i}\cdot \mathbf{v}(m)$, whereas the expected payoff of the co-players is given by $\pi_{-i}(m)=\mathbf{g_{-i}}\cdot \mathbf{v}(m).$ Finally, let $\mathbf{1}$ be the $2n$-dimensional vector with entries 1, and let $\mathbf{g_0}$ be the $2n$-dimensional vector in which the first $n$ entries are 1 and the last $n$ entries are 0, 
\begin{equation}
\begin{array}{ccl}
\mathbf{1}	&=	&\big(1,\ldots,1,1,\ldots,1\big)\\[0.2cm]
\mathbf{g_0}	&=	&\big(1\ldots,1,0,\ldots,0\big).
\end{array}
\end{equation}
We note that zero-determinant strategies are exactly those strategies for which
\begin{equation}
\mathbf{p}=\mathbf{g_0}+\phi\Big[(1-s)(l\mathbf{1}-\mathbf{g_i})+\mathbf{g_i}-\mathbf{g_{-i}}\Big].
\end{equation}
Finally, let $q_C(m)$ denote the probability that player $i$ cooperates in round $m$. We can write $q_C(m)=\mathbf{g_0}\cdot \mathbf{v}(m)$ and $q_C(m+1)=p\cdot \mathbf{v}(m)$. Thus, it follows for $w(m):=q_C(m+1)-q_C(m)$ that
\begin{equation} \label{Eq:w(m)} \small
w(m)=(\mathbf{p}-\mathbf{g_0})\cdot \mathbf{v}(m) = \phi\Big[(1-s)(l\mathbf{1}-\mathbf{g_i})+\mathbf{g_i}-\mathbf{g_{-i}}\Big]\cdot \mathbf{v}(m)=\phi\Big[s\pi_i(m)+(1-s)l-\pi_{-i}(m)\Big].\end{equation}
Taking the definition of $w(m)$, summing up over $m$, and taking the limit $M\to \infty$ yields
\begin{equation} \small
\frac{1}{M} \sum_{m=1}^{M} w(m)=\frac{1}{M} \sum_{m=1}^{M} q_C(m+1)-q_C(m) = \frac{q_C(M+1)-q_C(0)}{M} = \frac{q_C(M+1)-p_0}{M}.
\end{equation}
On the other hand, due to \eqref{Eq:w(m)},
\begin{equation} \small
\frac{1}{M} \sum_{m=1}^{M} w(m)=\frac{\phi}{M} \sum_{m=1}^{M} s\pi_i(m)+(1-s)l-\pi_{-i}(m) = \phi\big[s\pi_i+(1-s)l-\pi_{-i}\big]
\end{equation}
As the two limits need to coincide, and as $|q_C(M+1)-p_0|\le 1$, the result follows.
\end{proof}
~

\begin{proof}[\bf Proof of Proposition \ref{Prop:charpayrelSI}]
Let us first show that the parameters $s$,$\phi$ of a zero-determinant strategy satisfy 
$\frac{1}{n-1}\le s \le 1$ and $\phi>0,$ and that for $s<1$ the baseline payoffs fulfill $b_0\le l \le a_{n-1}$. By the definition of zero-determinant strategies, the cooperation probabilities after mutual cooperation and mutual defection are given by
\begin{equation}
\begin{array}{ccc}
p_{C,n-1}	&=	&1+\phi(1-s)(l-a_{n-1})\\
p_{D,0}	&=	&\phi(1-s)(l-b_0).
\end{array}
\end{equation}
As these two entries need to be in the unit interval, it follows that 
\begin{equation} \label{Ineq:l}
\begin{array}{clc} 
\phi(1-s)(l-a_{n-1})	&\le &0\\
0	&\le 	&\phi(1-s)(l-b_0)	
\end{array}
\end{equation}
Adding up these two inequalities implies $\phi(1-s)(b_0-a_{n-1})\le 0$, and therefore 
\begin{equation}\label{Ineq:phis1}
\phi(1-s)\ge 0.
\end{equation}
Analogously, by taking into account that the entries $p_{C,n-2}$ and $p_{D,n-1}$ need to be in the unit interval, one finds that
\begin{equation}\label{Ineq:phis2}
\phi\left(s+\frac{1}{n-1}\right)\ge 0.
\end{equation}
Adding up the two inequalities [\ref{Ineq:phis1}] and [\ref{Ineq:phis2}] shows that $\phi>0$. As a consequence, it follows from  [\ref{Ineq:phis1}] and [\ref{Ineq:phis2}] that $-1/(n-1)\le s \le 1$. Lastly, Inequalities (\ref{Ineq:l}) then imply for $s\neq 1$ that $b_0\le l \le a_{n-1}$.

Let us now turn to the characterization of enforceable payoff relations: If $s=1$, the entries of a zero-determinant strategy $\mathbf{p}$ according to representation [\ref{eq:defzdSI}] do not depend on the parameter $l$. By choosing $\phi>0$ sufficiently small, the inequalities $p_{S,j}$ are satisfied for any social dilemma (in fact, for any game in which $b_j\ge a_{j-1}$ for all $j$). 

Now suppose $s\neq 1$ and $\mathbf{p}$ is a zero-determinant strategy. As $\phi>0$,  the representation of zero-determinant strategies [\ref{eq:defzdSI}] implies the following constraints on the entries $p_{S,j}$:
\begin{equation} \label{Ineq:ExtZD}
\displaystyle
\begin{array}{clc}
(1-s)(l-a_j)-\frac{n-j-1}{n-1}(b_{j+1}-a_j)	&\le &0\\[0.4cm]
(1-s)(l-b_j)+\frac{j}{n-1}(b_j-a_{j-1})	&\ge &0.
\end{array}
\end{equation}
Dividing by $1-s>0$ shows that the inequalities [\ref{Ineq:ExtZD}] are equivalent to condition [\ref{eq:charrelSI}]. Conversely, suppose condition [\ref{eq:charrelSI}] and thus the inequalities [\ref{Ineq:ExtZD}] are met. Then only the parameter $\phi>0$ has to be chosen sufficiently small to ensure that all entries satisfy $0\le p_{S,j}\le 1$ in representation [\ref{eq:defzdSI}].
\end{proof}
~

\begin{proof}[\bf Proof of Proposition \ref{prop:allianceSI}]
Condition [\ref{eq:charallSI}] is derived from the corresponding condition [\ref{eq:charrelSI}] for solitary players, by using the relation between $s_\mathcal{A}$ and $s$ in Eq.~[\ref{eq:saSI}]. Thus, for symmetric alliances, the result follows directly from Proposition \ref{Prop:charpayrelSI}. It is only left to show that every linear relationship of the form [\ref{eq:zdallSI}] that can be enforced by an asymmetric alliance (in which allies may use different ZD strategies) can also be enforced by a symmetric alliance (in which all allies use the same ZD strategy). To this end, let us consider an asymmetric alliance $\mathcal{A}=\{1,\ldots,k\}$ that collectively enforces some linear relation between payoffs,
\begin{equation} \label{eq:zdall2SI}
\pi_{-\mathcal{A}}=s_\mathcal{A}\pi_\mathcal{A}+(1-s_\mathcal{A})l_\mathcal{A},
\end{equation}
and where each ally $i\in\mathcal{A}$ individually enforces the relation
\begin{equation} \label{eq:zdiSI}
\pi_{-i}=s_i\pi_i+(1-s_i)l_i.
\end{equation}
By summing up all equations [\ref{eq:zdiSI}] and rearranging the terms, we obtain
\begin{equation} \label{eq:zdassSI}
\pi_{-\mathcal{A}}=\sum_{i=1}^k\frac{s_i(n-1)-(k-1)}{n-k}\frac{\pi_i}{k}+\sum_{i=1}^k \left(1-\frac{s_i(n-1)-(k-1)}{n-k}\right)\frac{l_i}{k}
\end{equation}
As the right hand sides of [\ref{eq:zdall2SI}] and [\ref{eq:zdassSI}] need to coincide, we have
\begin{equation} \label{eq:zdassSI} \small
s_\mathcal{A}\sum_{i=1}^k \frac{\pi_i}{k}+(1-s_\mathcal{A})l_\mathcal{A}=\sum_{i=1}^k\frac{s_i(n-1)-(k-1)}{n-k}\frac{\pi_i}{k}+\sum_{i=1}^k \left(1-\frac{s_i(n-1)-(k-1)}{n-k}\right)\frac{l_i}{k}
\end{equation}
Both sides of this equation represent a linear function in the variables $\pi_i$. A comparison of coefficients yields 
\begin{equation} \label{eq:zdassSI} \small
s_\mathcal{A}=\frac{s_i(n-1)-(k-1)}{n-k}.
\end{equation}
In particular, it follows that $s_1=s_2=\ldots=s_k=:s$, i.e. an alliance of ZD strategists that enforces a linear relation between payoffs needs to coordinate on the same slope. In that case, $l_\mathcal{A}$ can be calculated as $l_\mathcal{A}=\sum_{i=1}^k l_i/k$, i.e. $l_\mathcal{A}$ is the arithmetic mean of the individual baseline payoffs $l_i$. In particular, $\min l_i \le l_\mathcal{A} \le \max l_i$. As all players of the alliance applied a ZD strategy, it follows that ($\min l_i,s$) and ($\max l_i,s$) are enforceable, and therefore also ($l_\mathcal{A}$,s) (as outlined in the first sentence after Proposition \ref{Prop:charpayrelSI}). Thus, there is a symmetric alliance, in which all allies use the same ZD strategy with parameters ($l_\mathcal{A},s$), that enforces the relation [\ref{eq:zdall2SI}]. 
\end{proof}
~

\begin{proof}[\bf Proof of Proposition \ref{Prop:nashSI}]
We already know that for a ZD-strategy to be a Nash equilibrium, one of the three conditions need to be fulfilled (otherwise there would be a different ZD-strategy that yields a higher payoff). Conversely, let us assume that one of the three conditions of the Proposition is fulfilled.
\begin{enumerate}
\item If $s_\mathcal{A}=0$, then the remaining player obtains a payoff of $\pi_{n}=l$, irrespective of $n$'s strategy. In particular, there is no incentive for player $n$ to deviate.
\item Suppose $s_\mathcal{A}>0$, $l=a_{n-1}$, and let us assume to the contrary that the zero-determinant strategy is not a Nash-equilibrium. Then there is a strategy for player $n$ such that $n$'s payoff satisfies $\pi_n>a_{n-1}$. However, as $\mathcal{A}$ enforces the relation $
\pi_n=s_\mathcal{A}\pi_\mathcal{A}+(1-s_\mathcal{A})a_{n-1}$,
and as $s_\mathcal{A}>0$, we can conclude that $\pi_\mathcal{A}>a_{n-1}$. It follows that the average payoff of all group members exceeds $a_{n-1}$, contradicting the assumption that $a_{n-1}$ is the maximum average payoff per round.
\item Under the assumption that $b_0$ is the minimum average payoff per round, the case $s_{\mathcal{A}}<0$ and $l=b_0$ can be treated analogously to the previous case.
\end{enumerate}
\end{proof}

\end{document}